\documentclass[showkeys,letterpaper,floatfix,nobalancelastpage,twocolumn]{revtex4}
\usepackage{bm}% bold math
\usepackage{amsmath}
\usepackage{bm,braket}
\usepackage{color}
\usepackage{float}
\usepackage{graphicx,subfigure}
\usepackage{amssymb}
\usepackage{sidecap}
\usepackage{float} % new added to located the figure at the given position
\begin{document}

%\title{Quasimode expansion technique for metallic nano-resonators: effective %mode volumes, Purcell factors and photon propagators for  quantum plasmonics}

\title{Quasinormal mode theory and modelling of electron energy loss spectroscopy for plasmonic nanostructures  }

\author{Rong-Chun Ge}
\email{rchge@physics.queensu.ca}
\author{Stephen Hughes}
\email{shughes@physics.queensu.ca}
\affiliation{
Department of Physics, Engineering Physics and Astronomy, Queens University, Kingston, Ontario, Canada K7L 3N6}

\keywords{metal resonators, electron energy loss spectroscopy (EELS), quasinormal modes, local density of optical states (LDOS), Green function,  Purcell factor, nanoplasmonics}

\begin{abstract}

Understanding  light-matter interactions using localized surface plasmons (LSPs) is of fundamental interest in classical and quantum plasmonics and has a wide range of applications. In order to understand the spatial properties of LSPs, electron energy loss spectroscopy (EELS) is a common and powerful method of spatially resolving the extreme localized fields that can be obtained  with metal resonators.
However, modelling EELS for general shaped resonators presents a major challenge
in computational electrodynamics, requiring the full photon Green function as a function of two space points and frequency. Here 
we present an intuitive and computationally simple method
for computing EELS maps of plasmonic resonators using a  quasinormal mode (QNM) expansion technique. By separating the contribution of the QNM and the bulk material, we give closed-form analytical formulas for the plasmonic QNM contribution to  the EELS maps. We exemplify our technique for
a split ring resonator, a gold nanorod, and a nanorod dimer structure. 
 The method is accurate, intuitive, and gives orders of magnitude improvements over direct
 dipole simulations that numerically solve the full 3D Maxwell equations.
 We also show how the same QNM Green function can be used to obtain the Purcell factor (and projected local density of optical states) from quantum dipole emitters or two level atoms,  and we demonstrate how the spectral features differ in general to the EELS spectrum.
\end{abstract}

%\keywords{metal resonators, electron energy loss spectroscopy, quasinormal modes, local density of optical states, Green function,  Purcell factor, quantum plasmonics}

\maketitle

\section{Introduction}
Nanoplasmonics continues to receive substantial interest from various  fields  of research including  biology~\cite{Bio1,Bio2}, chemistry~\cite{Chem1,Chem2} and physics~\cite{Phys},
with applications ranging from renewable energy technology~\cite{Plasmonicsphotovoltaic} to homeland security by the sensitive identification of explosive material~\cite{Secu}. Due to the collective excitations of electrons at the surface of a metallic resonator coupling to  electromagnetic fields, the local density of optical states (LDOS) decreases rapidly for spatial positions away from the surface of the metal resonator; so-called ``hot spots'', where the local electric field can be enhanced by orders of magnitude in comparison to a bulk medium, are formed near the surface. The unique properties of localized surface plasmon (LSP) resonances, manifest in a strong confinement of electric field far below the diffraction limit and result in an exotic electromagnetic response that underlines the wide application of surface plasmons, especially in nanoscience and nanophotonics; however, the extreme spatial localization  of LSPs makes them experimentally challenging for direct detection of the spatial field distribution, e.g., by employing conventional spectroscopy techniques, since the spatial resolution is constrained by the diffraction limit. In addition,  optically dark modes have a vanishing dipole moment, and traditional  optical methods fail to  excite such modes.

Due to the strong frequency dispersion and losses in metals, which 
satisfy the  Kramers-Kronig relations, it is a very challenging problem to model the electromagnetic response of arbitrarily shaped metal resonators. Consequently, most  optical studies of plasmonic structures  rely heavily on brute force numerical simulations. Direct imaging of the LSP resonances is important both for revealing the exotic physics underlying these resonances and for verifying and testing numerical models used in theoretical studies. There are presently several different experimental schemes developed to detect the electromagnetic component of LSPs, including near-field scanning optical microscopy (NSOM)~\cite{NSOM,NSOM1}, leakage radiation microscopy~\cite{LMS}, and two-photon induced photoluminescence~\cite{NTPL,TPL}, and cathodoluminescence~\cite{Los}.  Of particular interest, electron energy loss spectroscopy (EELS) has been shown to be capable of accessing the subwavelength spatial variation of the surface plasmon modes of single metallic nanoresonator experimentally~\cite{Nelayah,Bosman} for almost a decade; during this time, it has been recognized to be ~\cite{EELStheo,EELStheo2,EELSr} one of the most powerful approaches, in which high speed electron beams (typically ranging from several tens  to several hundreds of keV) are injected and transmitted across an optically  thin sample, with a spatial resolution of around 1~nm; EELS has been applied to various systems such as split ring resonator (SRR)~\cite{Husnik} and single nanorod/antennas~\cite{Nico,Ross}. The EELS method is   %\textbackslas\
 also quite  versatile, allowing one to probe both optically bright and dark resonances over broadband frequencies, and can be used for detecting both localized and extended excitations~\cite{EELSr}.
  Recently, vortex electron beams have also been experimentally 
  demonstrated~\cite{Vortexelectron,Vortexe2}, which may find use for
  probing the magnetic component of LSPs~\cite{VEELSGordon}.

Modelling EELS is an extremely challenging and tedious numerical problem, and  
there has been different theoretical approaches developed to model the EELS of plasmonic resonators, including boundary element method (BEM)~\cite{Ouyang,Abajo,Boundary,hohens}, discrete dipole approximation (DDA)~\cite{Henrard}, discontinuous Galerkin time-domain (DGTD)~\cite{DGTD,DGTD2} and finite-difference time-domain  (FDTD) methods~\cite{VEELSGordon,EELSCao}.
 %\blue{Add some refs and sentences to Hohenester and García de Abajo - what methods do they use and are they restricted to certain shapes}
 With sufficient care and computational resources, all of these approaches can show good agreement with experimental results, since these are basically full dipole solutions
to the classical Maxwell equations; typically, these approaches  employ numerous dipole point calculations in a 3D spatial grid to obtain the photon Green function, which allows one to obtain the EELS profile in space and frequency; non-local effect of the conductive electrons are typically negligible until spatial positions within  a few nm from the metal 
surface~\cite{Quantumplasmon1,Quantumplasmon2,Quantumplasmon3,Quantumplasmon4},  so most studies of EELS in nanoplasmonics  have worked with the classical Maxwell equations without any non-local effects; however, as shown very recently,
non-local effects may become important for certain metallic resonators and geometries~\cite{Christ,AsgerEELS}.
 Unfortunately, most  EELS calculation methods to date are  computational expensive, hard to employ without parallel computers, limited to certain geometries, and offer little physical insight into the LSP resonance structures. 
It is thus highly desired to find an approach that is both simpler
and more intuitive in terms of explaining the features of the EELS maps
and frequency profiles, and applicable to arbitrarily shaped resonators. Recently, the physical meaning and applications of EELS has been explored extensively. For example, H\"{o}rl {\it et al}.~\cite{Horl} show that  EELS is an efficient tomography probe of the surface plasmon modes, and they propose to get the 3D Green function from the EELS; 
unlike EELS, which is related to the projected full electromagnetic LDOS (ELDOS), Losquin {\it et al}.~\cite{Los} theoretically show that cathodoluminescence is related to the so-called projected radiative ELDOS (i.e., nonradiative coupling effects are not captured), which they illustrate with a  quasistatic  mode expansion technique based on the  BEM.  The quasistatic BEM approach uses only geometry-defined modes, which gives a nonretarded modal solution with scale invariance~\cite{Bouda}; in contrast,   the mode expansion technique  introduced below  uses the rigorously defined QNMs which are the true open system retarded eigenmodes, and we show only a few of them (indeed usually only one) will be needed around the frequency of interest.

\begin{figure}[t]
%\centering\includegraphics[trim=0cm 0cm 0cm 0cm, clip=true,width=.9\columnwidth]{drawing-SRR1.eps}
\centering\includegraphics[trim=0cm 0cm 0cm 0cm, clip=true,width=.99\columnwidth]{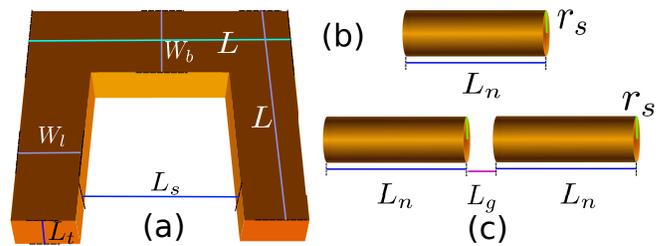}
\caption{
Schematic showing a selection of metallic nanoresonators: (a) split ring resonator; (b) nanorod; (c) dimer of two nanorods. The scale of the systems are $L = 200~$nm, $W_b = 80~$nm, $W_l = 50~$nm, $L_t = 30~$nm, $L_s = 100~$nm, $L_n = 100~$nm, $L_g = 20~$nm and $r_s = 15~$nm, and we consider material properties for gold. }
\label{f0}
\end{figure}

In this work, we introduce an accurate and physical intuitive method to model the EELS map of the surface plasmon modes based on the quasinormal mode (QNM) expansion  of the photon Green function~\cite{quasi,quasi2}. 
The QNMs are the eigenfunction of the source-free Maxwell equations with open boundary condition~\cite{Lai,Leung,Acspho}, and complex eigenfrequencies. %\blue{Define formally what a QNM is and give some refs to Lee and the ACS Photonics Paper}
Two key advantages of our QNM technique are as follows: (i) after obtaining the normalized QNM, the calculation of the EELS is straightforward and essentially instantaneous in the frequency regime of interest; (ii) our calculation includes
the contribution of the LSP in a modal theory, and thus has intuitive and analytical insight. 
After introducing the basic theory of EELS and connecting to the QNMs, we present several example structures of interest including a gold SRR, a single gold nanorod, and a dimer of gold nanorods as is shown schematically in Fig.~\ref{f0}. We also use the same QNM Green function to obtain the Purcell factor (and projected LDOS) from a coupled dipole emitter and we show how the spectral profile compares and contrasts with the EELS as a function of frequency. 

\section{Photon Green function expressed in terms of the QNMs}
%We then show the our method could be easily employed to get the VEELS too, %by using a QNM expansion for the magnetic Green function.

For the structures of interest, we
consider a general shaped metallic nanoresonator inside a homogeneous background medium with refractive index $n_B$. We assume the magnetic response is negligible, with permeability $\mu=1$;  the electric response is described by the Drude model, with permittivity $\varepsilon(\omega) = 1-\omega_p^2/(\omega^2+{\rm i}\omega\gamma)$ with parameters similar to gold: $\omega_p/2\pi = 1.26\times10^{16}~$THz and $\gamma/2\pi = 1.41\times10^{14}~$THz. The electric-field Green function, {\bf G}, of the system is defined as 
\begin{align}
 \nabla\times\nabla\times{\bf G}({\bf r},{\bf r}';\omega)
&-\frac{\omega^2}{c^2}\varepsilon({\bf r},\omega)
{\bf G}({\bf r},{\bf r}';\omega)\nonumber \\  
& =\frac{\omega^2}{c^2}{\bf I}\delta({\bf r}-{\bf r}'),
\end{align}
where ${\bf I}$ is the unit dyadic, and $\varepsilon({\bf r},\omega) = \varepsilon(\omega)$ inside the metallic nanoresonator with $\varepsilon({\bf r},\omega) = \varepsilon_{\rm B} = n_{\rm B}^2$ elsewhere. Considering the frequency regime of interest where there is only a single QNM, $\tilde{\bf f}_{\rm c}$, which gives the mode profile of the lossy/dissipative mode of the source free Maxwell equations with open boundary conditions, with complex eigenfrequency $\tilde{\omega}_{\rm c}$, then the contribution to the transverse Green function in the near field of the nanoresonator, around the cavity resonance,  is given by~\cite{quasi}
\begin{align}
{\bf G}^{\rm c}({\bf r},{\bf r}';\omega) = \frac{\omega^2\tilde{\bf f}_{\rm c}({\bf r})\tilde{\bf f}_{\rm c}({\bf r}')}{2\tilde{\omega}_{\rm c}(\tilde{\omega}_{\rm c}-\omega)}.
\label{GT}
\end{align}
The QNM, $\tilde{\bf f}_{\rm c}({\bf r})$ is normalized here as $\langle\langle \tilde{\bf f}_{\rm c}|\tilde{\bf f}_{\rm c}\rangle\rangle\!=\!\lim_{V\rightarrow\infty}\int_V\sigma({\bf r},\omega) \,\,\,\!\tilde{\bf f}_{\rm c}({\bf r})\cdot\tilde{\bf f}_{\rm c}({\bf r})d{\bf r} + \frac{ic}{2\tilde{\omega}_{\rm c}}\int_{\partial V}\sqrt{\epsilon({\bf r})}\tilde{\bf f}_{c\rm }({\bf r})\cdot\tilde{\bf f}_{\rm c}({\bf r})d{\bf r}=1$ with $\sigma({\bf r},\omega) = \partial(\varepsilon({\bf r},\omega)\omega^2)/2\omega\partial\omega|_{\omega = \tilde{\omega}_{\rm c}}$. Alternative QNM normalization schemes are discussed 
in~\cite{Sauvan,Philip}.

\section{Effective mode volume and Purcell factor}
When discussing  EELS, it is useful to also connect to common quantities for use in quantum plasmonics. For example, using the normalized QNM, the corresponding effective mode volume for use in Purcell factor calculations
is defined as
\begin{align}
V_{\rm eff}({\bf r}_0) = 1/{\rm Re}
\left[\frac{\varepsilon_B\tilde{\bf f}_{\rm c}^2({\bf r}_0)}{
\langle\langle\tilde{\bf f}_{\rm c}|\tilde{\bf f}_{\rm c}\rangle\rangle}\right],
\label{eq:Veee}
\end{align}
 at some characteristic position ${\bf r}_0$~\cite{Philip}.  
 The enhancement of spontaneous emission (SE), or the 
  enhancement of the projected LDOS,
 at this position is  then obtained from
\begin{align}
F_{\alpha}({\bf r}_0,\omega) = \frac{{\rm Im}[\hat{n}_\alpha\cdot{\bf G}({\bf r}_0,{\bf r}_0;\omega)\cdot\hat{n}_\alpha]}{{\rm Im}[\hat{n}_\alpha\cdot{\bf G}_{\rm B}({\bf r}_0,{\bf r}_0;\omega)\cdot\hat{n}_\alpha]},
\label{eq:PF}
\end{align}
where ${\rm Im}[{\bf G}_{\rm B}({\bf r},{\bf r};\omega)] = \frac{\omega^3n_{\rm B}}{6\pi c^3}{\bf I}$
 is for a lossless homogeneous background with refractive index $n_{\rm B}$, and  $\hat{n}_\alpha$ is a unit vector of the dipole emitter aligned along
 $\alpha = x,y,z$. Using the QNM approach, then $F_\alpha$ is simply obtained by using
 ${\bf G}\approx{\bf G}^{\rm c}$, and the accuracy of this approach can be checked by performing a full dipole calculation of ${\bf G}$ at this position, which we will show later using  
accurate FDTD techniques \cite{FDTD}.

\section{Theory and modelling of EELS} 
We consider a high speed electron beam with initial kinetic energy, $E_k$, e.g., 50 keV$\sim$200~keV, which gives an electron speed, $|{\bf v}|$, 0.55\,$c\sim$ 0.70 $c$ with $c$ the velocity of light in vacuum; specifically, $|{\bf v}| = c\sqrt{1-\frac{m_e^2c^4}{(E_k+m_ec^2)^2}}=0.55c$ with $m_e$ the electron rest mass; the electron passes through the nanoresonator which is a few tens of nanometers thick. Generally, scanning transmission electron microscopes will be employed  to obtain the EELS map, as a function of frequency, for which the relevant length scale of the spatial path over which the  electron beam is traveling is around a few hundred~nanometers;
 %\blue{not clear what you mean here - the beam of course is much longer than that?},
under this situation, the energy loss of the electron is negligible, which means that, to  a very good approximation, we can take the velocity of the electron as a constant. 
As the electron comes near the surface, the electric quasistatic interaction can be described by the image charge~\cite{quasi,image}, which is negligible until it comes to around a few nanometers from the surface; at this scale, the local geometry details can be ignored and the surface can be approximated by a slab, and detailed analysis elsewhere shows that the electric quasistatic contribution is typically negligible~\cite{note} for the EELS calculation. Thus we assume the electron energy loss is primarily induced by the dominant QNM(s). 
%\blue{This paragraph needs to be shortened and made clearer, possibly adding some stuff to a ref/note}
In practice there are also ``bulk losses''\cite{EELSr} coming from other background modes such as evanescent modes in the metal, but these are regularized depending upon the finite size of the cross section of the electron beam and have little influence on a modal interpretation of the EELS map. In fact, in~\cite{EELSCao},  in  order to investigate the  modal response of the LSP resonance of the nanoresonator, they eliminated the bulk contribution by subtracting the solution from a different   FDTD simulation with a homogeneous metal calculation, thus eliminating  FDTD grid-dependent effects. With our QNM approach, there is no need to subtract off  such a term, and, moreover, this contribution can be obtained analytically 
\cite{EELSr}, and can also formulated 
as a local field problem for emitters inside lossy resonators~\cite{OpticaQNM}. Numerically, we inject a spatial plane wave modulated with a finite pulse length (FWHM), $\Delta t$, with a central frequency around the resonance of the QNM; then a run-time Fourier transform with a time window $\Delta\tau$  is employed~\cite{quasi2} to get the QNM numerically; we also use a non-uniform conformal mesh scheme with a fine mesh of  1 - 2~nm around the metallic nanoresonator.

The energy loss is defined by %\blue{define all terms, such as $v$ and $E^{in}$}
\begin{align}
\Delta E = \int e{\bf v}(t)\cdot{\bf E}^{\rm in}({\bf r}_t,t)dt =\int_0^\infty\hbar\omega\Gamma(\omega)d\omega,
\end{align}
where the electric field induced by the QNM is given by 
\begin{align}
&{\bf E}^{\rm in}
({\bf r}_t,t)\nonumber\\
= &\int_{-\infty}^{\infty}{\bf E}^{\rm in}({\bf r}_t,\omega)e^{-{\rm i}\omega t}d\omega\nonumber\\
=&2\int_0^{\infty}{\rm Re}[{\bf E}^{\rm in}({\bf r}_t,\omega)e^{-{\rm i}\omega t}]d\omega\nonumber\\
=&-2\int_{0}^{\infty}d\omega {\rm Im}[e^{-{\rm i}\omega t}\frac{1}{\varepsilon_0\omega}\int{\bf G}^{\rm c}({\bf r}_t,{\bf r}';\omega)\cdot{\bf j}({\bf r}',\omega)d{\bf r}']
\end{align} 
with the effective current carried by the moving electron, ${\bf j}({\bf r},\omega) = \frac{1}{2\pi}\int{\bf j}({\bf r},t')e^{{\rm i}\omega t'}dt'
  =-\frac{e}{2\pi}\int{\bf v}\delta(r'-r_{t'})e^{{\rm i}\omega t'}dt'$, and $e$ is the absolute value of the charge of the electron. In the  calculations below, we will assume the electron moving along the -$z$-axis, so ${\bf v} = -\hat{n}_zv$ with $\hat{n}_z$  the unit vector along $z$. 
Under these assumption, the EELS function, $\Gamma(\omega)$, due to the QNM for electrons injected along $z$-axis is simply given by~\cite{EELStheo} %\blue{not clear to me where this comes from, so show a few more steps and give $E^{in}$ expression and how it relates to G}
  \begin{align}
  &\Gamma(\omega,R_0) = \nonumber \\
  & \ \ \frac{e^2}{\hbar\varepsilon_0\omega^2\pi}\int\int{\rm Im}[{\rm G}^{\rm c }_{zz}(R_0,z,z';\omega)e^{{\rm i}\omega (z'-z)/v}]dzdz', 
%  \frac{e^2v^2}{\hbar\varepsilon_0\omega^2\pi}\int\int{\rm Im}[{\rm G}^{c}_{zz}({\bf %r}_t,{\bf r}_{t'};\omega)e^{{\rm i}\omega (t'-t)}]dtdt', \nonumber\\
%  =&\frac{e^2}{\hbar\varepsilon_0\omega^2\pi}\int\int{\rm Im}[{\rm G}^{c %}_{zz}({\bf r}_t,{\bf r}_{t'};\omega)e^{{\rm i}\omega (z'-z)/v}]dzdz', 
  \label{LOSS}
  \end{align} 
where $G^{\rm c}_{zz} = \hat{n}_z\cdot{\bf G}^{\rm c}\cdot\hat{n}_z$ and $R_0=(x,y)$ on a 2D spatial map of the image. As is shown in Eq.~(\ref{LOSS}), in order to calculate the EELS for one particular $R_0$ in the plane, the Green function along the electron beam should be calculated at various $z$
and $z'$; the number of simulations required should be sufficiently large to model electric dipoles scanning over the trajectory of the electron beam, and this is the reason why thousands of dipole  simulation are usually employed~\cite{EELSCao}. In stark contrast, with the QNM technique, once the QNMs are obtained numerically, the Green function can be calculated with Eq.~(\ref{GT}) analytically, with a computation that is basically instantaneous.

\section{numerical Results and example calculations for various metal resonators}
Below we   present a selection of example metal resonators using the  Drude model for the material properties of the metal.
We also show the Purcell factor or enhanced SE factor at selected positions as well as the full EELS as a 2D image. 

\subsection{Split ring resonator} 

\begin{figure}
%\centering\includegraphics[trim=0.1cm 2.2cm 0.18cm 2.5cm, clip=true,width=.98\columnwidth]{SRRtry.eps}
\centering\includegraphics[trim=3.0cm 0.8cm 1.0cm 0.0cm, clip=true,width=.98\columnwidth]{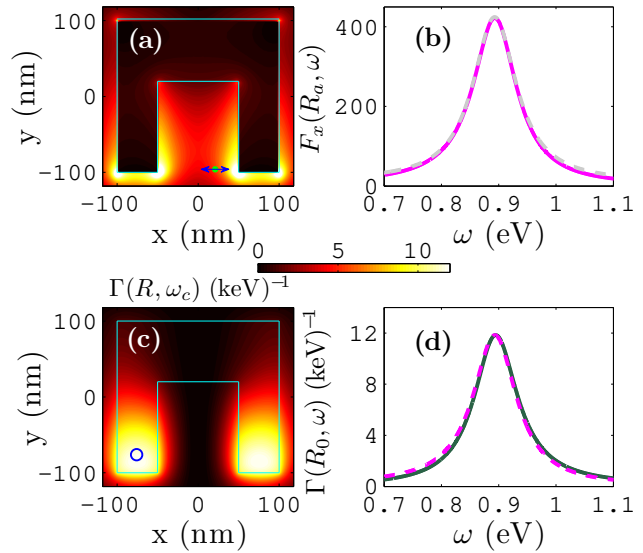}
\caption{EELS and SE enhancement for the SRR. (a) QNM profile $|\tilde{\bf f}(x,y,0;\omega_{\rm c})|$ with $\omega_{\rm c}/2\pi=\omega_1/2\pi = 216.16~$THz. (b) Enhanced SE factor, $F_x$ [Eq.~(\ref{eq:PF})], calculated using the QNM Green function (magenta solid), Eq.~(\ref{GT}), and full numerical FDTD calculation (grey dashed) for a $x$-polarized dipole at position ${\bf r}_a = (21,-96, -3)~$nm; see arrow shown in (a). (c) The spatial distribution of the EELS for injected electrons with energy $E_k=100$~eV ($v = 0.55c$) along the $z$-axis at $\omega_{\rm c}$;  (d) EELS (dark green solid) as a function of frequency at position, $R_0= ( 21,-96)~$nm, shown by the blue circle in (c); the magenta dashed is the  enhanced SE factor as is shown in (b). The cyan lines show the boundary of the SRR.}
\label{f1}
\end{figure}

For our first example of the QNM calculation of EELS, we study the gold SRR, which is the basic ``artificial atom'' unit cell of the negative index metamaterial, with  a rich  magnetic response to external electromagnetic fields~\cite{SRREELSex,DGTD2,VEELSGordon}. While the 2D metamaterial lattice is usually fabricated on a low index semiconductor, for the present study, we will ignore the effects of the substrate and assume the SRR is in free space ($n_B =1$), though  this is not a model restriction. The SRR with thickness, $L_t = 30~$nm,  is located in $xy$-plane as is shown in Fig.~\ref{f0}(a) with parameters $L = 200~$nm, $W_b = 80~$nm, $W_l = 50~$nm and $L_s = 100~$nm.

A full dipole FDTD calculation~\cite{quasi2,FDTD} shows that the dipole resonance of the QNM is around $\tilde{\omega}_1/2\pi = (\omega_1-{\rm i}\gamma_1)/2\pi = (216.16 - {\rm i} 11.15)~$THz with mode profile $|\tilde{\bf f}(x,y,0;\omega_{\rm c})|$ shown in Fig.~\ref{f1}(a). We use an $x$-polarized spatial plane wave with central frequency around 216~THz and pulse width $\Delta t=3$~fs, which is injected along the $z$-axis; a running Fourier transform with a temporal bandwidth~\cite{quasi2}  $\Delta\tau = 81~$fs is   used to obtain the QNM. The corresponding effective mode volume (which is  complex in its generalized form \cite{Philip}) is around $V_{\rm eff}({\bf r}_{\rm a})\approx 1.7\times 10^{-3}(\lambda_1/n_B)^3$ at the chosen dipole position
${\bf r}_{\rm a} = (23, -96, -3)~$nm where $\lambda_1 = 2\pi c/\omega_1$ (the dipole is shown by the blue arrow in Fig.~\ref{f1}(a), and note we have set the center of the SRR as the origin of the coordinate system). %\red{since the antinode should diverge at a sharp surface, I  think it may be better to use the dipole position shown on the figure, i.e., the $R_o$ position}. 
In order to first check the accuracy of the QNM calculation, we calculate the enhancement of the projected LDOS  (or SE enhancement of a dipole emitter) $F_{\alpha}({\bf r},\omega)$ from Eq.~(\ref{eq:PF}). Figure~\ref{f1}(b) shows that the single QNM model calculation (magenta solid, with Eq.~(\ref{GT})) agrees very well with the full numerical dipole calculation using 
FDTD~\cite{RaoPRL,YaoLaser,ColeOL} (grey dashed) at position ${\bf r}_{\rm a}$. {As is shown above, due to the strong confinement of the LSP, extremely small mode volumes are obtained leading to a strong enhancement of SE of an electric dipole (or single photon emitter) at this near field spatial position. The broad bandwidth of the SE enhancement is  also a notable feature of metallic nanoresonators, making it much easier to spectrally couple to artificial atoms. We further remark that the spatial dependent spectral function, $ {\rm Im}[{\bf G}({\bf r},{\bf r},\omega)]$, as is discussed in Ref.~\cite{Ourprb15,plasmonresonance}, usually has a non-Lorentzian lineshape that in general changes as a function of position; this effect is  captured by the  QNM technique~\cite{Ourprb15} through the spatial dependent phase factor of the QNM}. 

Due to the collective motion of the free electrons, there is an oscillating electric current circling along the SRR; as a result, a temporarily non-zero magnetic dipole is created, which displays a strong magnetic response to external electromagnetic field around the resonance of the QNM and forms the basis of exciting optical properties of metamaterial such as negative index. The working region %\red{what is meant by this?}
 of the SRR could be controlled by changing the length of the SRR which determines the QNM resonance.

The 2D EELS image, $\Gamma(R,\omega)$ for a high energy, $E_k = $100~keV, ($v = 0.55 c$) %\blue{why this value specifically, and e.g., how would it differ for 60, 60; and what is the corresponding
%$v$ and give formula for conversion }, 
electron beam injected along $z$-axis is shown in Fig.~\ref{f1}(c), with   $\omega_{\rm c} = \omega_1$, which is consistent with the calculations in Refs.~\cite{SRREELSex,DGTD2,VEELSGordon} using full FDTD, nodal DGTG, and BEM, respectively. The EELS at the position of the blue circle
(near a maximum field position) is obtained as shown in Fig.~\ref{f1}(d) by the dark green solid line as a function of the frequency;  clearly the EELS can be used to effectively explore the QNM response
of the plasmonic resonator. {The magenta dashed line in Fig.~\ref{f1}(d)  shows that   the SE enhancement example  almost has the same spectral lineshape as the spatially averaged EELS calculation.}
%\red{comment on how well the PF lineshape and the EELS one agrees for this case}

\subsection{Gold nanorod} 

\begin{figure}[b]
%\centering\includegraphics[trim=1.1cm 0.9cm 1.0cm 3.4cm, clip=true,width=.98\columnwidth]{rEELS_nano100keV.eps}%{nanorod60kev.eps}
%\centering\includegraphics[trim=1.1cm 0.9cm 1.1cm 3.4cm, clip=true,width=.98\columnwidth]{nanotryr.eps}%{nanorod60kev.eps}
\centering\includegraphics[trim=0.5cm 0.1cm 4.1cm 0.0cm, clip=true,width=.98\columnwidth]{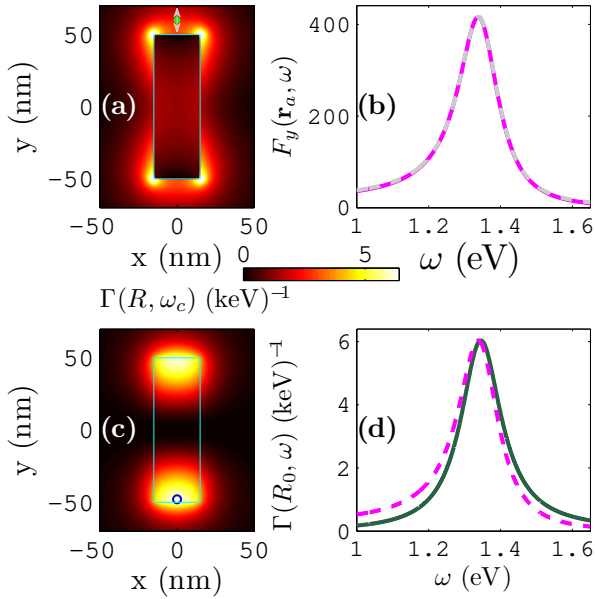}
%\centering\includegraphics[trim=1.1cm 0.9cm 1.0cm 3.4cm, clip=true,width=.98\columnwidth]{nanorod60kev.eps}
\caption{EELS and SE enhancement for the nanorod. (a) QNM profile $|\tilde{\bf f}(x,y,0;\omega_{\rm c})| $ at $\omega_{\rm c}/2\pi=\omega_2/2\pi = 324.98~$THz. (b) QNM (magenta solid) and FDTD (light dashed) calculation of $F_y({\bf r}_{\rm a},\omega)$ as a function of frequency at ${\bf r}_a = (0,60,0)~$nm as is shown by the arrow in (a). (c) Spatial distribution of the EELS, $\Gamma(R,\omega_{\rm c})$, for injected electrons with energy $E_k = 100$~keV along $z$-axis. (d) $\Gamma(R_0,\omega_{\rm c})$ (dark green) at position $R_0 = (0,-48)~$nm as is shown by the blue circle in (a); the magenta dashed is the scaled enhancement of SE as is shown in (b). The cyan lines show the cross section of the nanorod}
\label{f2}
\end{figure}

\begin{figure}[t]
%\centering\includegraphics[trim=1.1cm 0.9cm 1.0cm 3.6cm, clip=true,width=.98\columnwidth]{EELS_dimer2.eps}
%\centering\includegraphics[trim=1.1cm 0.9cm 1.0cm 3.6cm, clip=true,width=.98\columnwidth]{rEELS_dimer100keV.eps}
%\centering\includegraphics[trim=1.7cm 0.9cm 0.5cm 2.6cm, clip=true,width=.98\columnwidth]{dimertryr.eps}
\centering\includegraphics[trim=2.6cm 0.1cm 1.5cm 0.0cm, clip=true,width=.98\columnwidth]{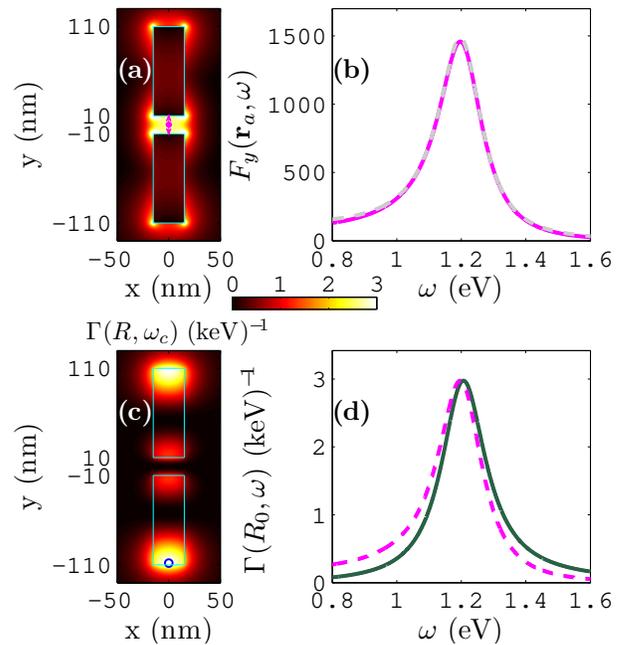}
\caption{EELS and SE enhancement factor for the gold dimer. (a) QNM profile $|\tilde{\bf f}(x,y,0;\omega_{\rm c})| $ at $\omega_{\rm c}/2\pi=\omega_3/2\pi = 291.06~$THz. (b) QNM (magenta solid) and full numerical calculation of the $F_y({\bf r}_{\rm a},\omega)$ at ${\bf r}_{\rm a} = (0,0,0)~$nm as shown by the arrow in (a). (c) Spatial distribution of $\Gamma(R,\omega_{\rm c})$ for injected electrons with energy $E_k=100$~keV along $z$-axis. (d) $\Gamma(R_0,\omega)$ as a function of frequency at $R_0 = (0, -108)~$nm as is shown by the blue circle; the magenta dashed is the scaled enhancement of SE as is shown in (b). The cyan lines show the cross section of the dimer.}
\label{f4}
\end{figure}

For our second example, we consider a single nanorod as is shown in Fig.~\ref{f0}(b), with radius ${\bf r}_s = 15~$nm and length $L_n=L/2 = 100~$nm. Frequently such nanorods are embedded inside  liquids, so  we assume the nanorod is located inside a homogeneous background medium with $n_{\rm B} = 1.5$. %\blue{say why you now use this}. 
Using FDTD simulations, 
the dipole resonance is found around $\tilde{\omega}_2/2\pi = (\omega_2-i\gamma_2)/2\pi (324.98-{ i}16.58)~$THz~\cite{quasi}. In order to get the QNM, a $y$-polarized spatial plane wave with central frequency 325~THz and $\Delta t = 6~$fs is injected along $x$-direction, and the running Fourier transform with $\Delta\tau = 60~$fs is employed~\cite{Ourprb15}. Figure~\ref{f2}(a) shows the mode profile of $|\tilde{\bf f}(x,y,0;\omega_2)| $ and the effective mode volume is found around $V_{\rm eff}({\bf r}_{\rm a}) \approx 1.8\times 10^{-3}(\lambda_c/n_{\rm B})^3$;  %\red{same comment as before, give at $R_0$}. 
the enhancement of the projected LDOS at position ${\bf r}_a = (0,60,0)~$nm  is shown in Fig.~\ref{f2}(b) (see arrow for dipole position), and the QNM calculation (magenta solid) shows excellent agreement with the full numerical calculation (grey dashed).  %QNM has been shown working excellent elsewhere~\cite{quasi}
%\blue{not sure for teh reader what you mean here}. 
The corresponding 2D EELS for an injected electron beam with energy $E_k = 100~$keV is shown in Fig.~\ref{f2}(c) at the resonance frequency, $\omega_{\rm c}/2\pi=\omega_2/2\pi = 324.98~$THz.  At the in-plane position $R = R_0 = (0,-48)~$nm, around which the maximum of the EELS is obtained (blue circle in Fig.~\ref{f2}(c)), $\Gamma(R_0,\omega)$ is shown by the blue solid line in Fig.~\ref{f2}(d). It can be seen that the EELS again picks up the correct resonant  response of the QNM. 
As is shown in Fig.~\ref{f4}(d), by the magenta dashed line, the SE enhancement
now has a different lineshape than the EELS calculation; as discussed earlier, this is caused by the spatially varying nature of spectral lineshape.

\subsection{Gold nanorod dimer}

For our final  resonator example, we study a dimer composed of two identical gold nanorods in homogeneous background with ${\bf n}_{\rm B} = 1.5$ as is shown in Fig.~\ref{f0}(c). The eigenfrequency of the dipole mode is found at $\tilde{\omega}_3/2\pi = (\omega_3-i\gamma_3)/2\pi=(291.06-{i}20.28)~$THz~\cite{quasi2}, and the correspondent QNM, $|\tilde{\bf f}(x,y,0;\omega_3)$ is shown in Fig.~\ref{f4}(a). Here in order to obtain the QNM mode, a $y$-polarized plane wave with central frequency 291~THz and $\Delta t = 6~$fs is injected, and $\Delta\tau = 46.6~$fs is used for the running Fourier transform~\cite{quasi2}. The  effective mode volume at ${\bf r}_{\rm a} = (0,0,0)~$nm is found to be  $V_{\rm eff}({\bf r}_{\rm a}) = 3.7\times 10^{-4} (\lambda_3/n_B)^3$;  the enhancement of the LDOS, $F_y({\bf r}_{\rm a},\omega)$ at ${\bf r}_{\rm a}$, using QNM and full numerical FDTD calculations are shown by the magenta solid and gray dashed in Fig.~\ref{f4}(b), respectively.
The 2D EELS, $\Gamma(R,\omega_3)$ is shown in Fig.~\ref{f4}(c) at $\omega_3/2\pi = 291.06~$THz, for injected electrons with energy $E_k  = 100~$keV; $\Gamma(R_0,\omega)$ at $R_0 = (0,-108)~$nm (blue circle in Fig.~\ref{f4}(a)) is shown by the dark green solid line in Fig.~\ref{f4}(d); { once again we see  that the SE enhancement (magenta dashed line)  displays a rather different lineshape compared to EELS calculation, similar to the case of the single nanorod. The SE enhancement is also much larger for the dimer which also has a larger output coupling efficiency~\cite{quasi2}.}

As is shown in~\cite{Ourprb15}, due to the inherent loss of the plasmonic system, the normalized QNM is in general complex, $\tilde{\bf f}({\bf r}) = |\tilde{\bf f}({\bf r})|e^{i\theta({\bf r})}$, and one obtains a position dependent phase factor $\theta({\bf r})$, which induces both the reshaping of the spectrum of LDOS (that is proportional to the imaginary part of the Green function) and the variation of the position of peak; this causes the spectral lineshape between the EELS and the enhancement of LDOS to differ in general.

\section{Discussion}

It is important to stress that for the QNM calculation of EELS, one only needs two simulations to get the complex eigenfrequency and  spatial distribution of the QNM (or QNMs), respectively. The rest of the calculation can be done semi-analytically with Eq.~(\ref{GT}). Consequently, the QNM technique for EELS is many orders of magnitude faster than full FDTD calculation using electric dipoles, and offers more insight. 
Furthermore, 
by obtaining the magnetic field component of the QNM, i.e., $\tilde{\bf h}_{\rm c} = -\frac{\rm i}{\mu_0\tilde{\omega}_{\rm c}}\nabla\times\tilde{\bf f}_{\rm c}$, the magnetic Green function, ${\bf G}_H({\bf r},{\bf r}';\omega) = \frac{\omega^2\tilde{\bf h}_{\rm c}({\bf r})\tilde{\bf h}_{\rm c}({\bf r}')}{2\tilde{\omega}_{\rm c}({\tilde\omega_{\rm c}-\omega})}$  could be calculated just as easily as  the electric Green function; this can be used to  simulate vortex-EELS as done in Ref.~\cite{VEELSGordon} which used brute force FDTD simulations.  The QNM could also be applied to model the electromagnetic force on atoms and nanostructures for optical trapping, which as input, usually requires the Maxwell stress tensor and/or the Green function \cite{NovotnyBook}.

Recently, Guillaume {\it et al}.~\cite{Guil} proposed an efficient modal expansion DDA method to model EELS.
 They show that by choosing a small number, e.g., 3-10, of eigenvectors their eigenvector expansion technique could get good results with reasonable accuracy, and the number of usual DDA operations are decreased considerably.
For certain geometries this approach may outperform a QNM FDTD computation, though it is not clear how general the approach is for various shaped resonators.   The philosophy of the  QNM approach is to use the source free eigenmode solutions, obtained here using  FDTD with the aid of the modal response from a scattered plane wave~\cite{quasi2}.
As a result, when there are a few dominant QNMs around the frequency of interest, in principle a single FDTD simulation is enough to obtain the modes and Green function as long as the overlap between the modes and injected field is sufficient. Moreover, we stress that the QNM computation is not limited to the FDTD method, e.g., it can also be obtained using an efficient dipole excitation technique with COMSOL~\cite{Bai}.
The typical time needed for our QNM calculations with a fine mesh size as small as 1~nm around the metal resonator, and total simulation volume 1-2 micron cubed, takes around several days on a high performance workstation. This is certainly not insignificant, however, having the full Green function as a function of position and frequency can then solve numerous problems without any more numerical simulations for the electromagnetic response.

{Apart from an efficient calculation of EELS, the QNM Green functions that we use above can be immediately adopted to efficiently study quantum light-matter interactions and true regimes of quantum plasmonics with quantized fields.
For example, using  the quantization scheme of the electromagnetic field in lossy structures~\cite{StefanPRA58,Dung98,WelschPRA99,ourPRBMollow}, the interaction between a quantum dipole (two level atom with frequency $\omega_0$ and dipole ${\bf d}$) and electric field operator in the rotating wave approximation (assuming no  external field) is given by the interaction Hamiltonian $H_{\rm I}=-\left[\sigma^+e^{i\omega_0t}\int_0^{\infty}d\omega\,{\bf d}\cdot \hat{\bf E}({\bf r}_k,\omega)+{\rm H.c.}\right]$; where the electric field operator is $\hat{\bf E}({\bf r},\omega)={ i}\int d{\bf r}'{\bf G}({\bf r},{\bf r}';\omega)\cdot
\sqrt{\frac{\hbar\varepsilon_I({\bf r}',\omega)}{\varepsilon_0\pi}}\hat{\bf f}({\bf r'},\omega)$, with $\varepsilon_I({\bf r},\omega)$ the imaginary part of $\varepsilon({\bf r},\omega)$ and $\hat{\bf f}({\bf r'},\omega)$ is the collective excitation operator of the field and medium; and $\sigma^+$ is the Pauli operator. we stress that the electromagnetic response of the lossy structure is rigorously included by the classical Green function (obtained from Eq.~(\ref{GT})).
For example, in a Born-Markov approximation,
 the quantum dynamics of $N$ quantum emitters around a metal nanostructures can be described
 through a reduced density matrix whose coupling terms can be fully described, including emitter-emitter and emitter-LSP interactions, through the analytical properties of the QNM Green function. For example, a recent example study of  the quantum dynamics between two plasmon-coupled quantum dots is shown in~\cite{Ourprb15}.
Note that such an approach is ultimately more powerful than a standard Jaynes-Cummings model since it can include non-Lorentzian decay processes and nonradiative coupling to the resonator in a self-consistent way, and it can also be used to improve the simpler Jaynes-Cummings models (in a regime where they are deemed to be approximately valid) with a rigorous definition of the various required coupling parameters~\cite{Ourprb15}.

} 
%\red{Extend this now with some discussions about how the same G, probed in EELS, is naturally used in quantum optics - perhaps with a few equations, showing how the E-field operators, ME, and spectrum (but not too much) all use the G - ref the two dot paper as an example - stress that such an approach is more accurate and can be used with a JC model - add refs as appropriate - no more than a page, or half a page is fine}

\section{Conclusions}
We have introduced   an efficient and  semi-analytic calculation  technique for modelling EELS using a QNM expansion technique, and exemplified the approach for several different metallic nanostructures. We first showed that the QNM technique works well for the SRR, and demonstrated that the QNM could be used to obtain similar 2D EELS maps to those shown in Refs.~\cite{SRREELSex,DGTD2,VEELSGordon}, but with orders of magnitude improvements in  efficiency and deeper physical insight. We then showed QNM calculations for a single gold nanorod and dimer of gold nanorods. We also presented example Purcell factor calculations and demonstrated how the spectral profiles may differ to EELS.

\acknowledgements
This work was supported by
the Natural Sciences and Engineering  Research Council of Canada
and Queen's University. We thank Philip Kristensen
for useful discussions.

%\com{add our usual acknowledgements}
%This work was supported by the Natural Sciences and
%Engineering Research Council of Canada. 

%We thank
%Jeff Young and
%Philip Kristensen for useful discussions.

%\pagebreak


\begin{thebibliography}{99}
\bibitem{note}
The contribution of image charge when the electron coming to the slab is to accelerate the electron, while it will decelerate the electron when the electron is leaving the other side of the slab. Both process will be antisymmetric of each other since we have assume the speed change of the process is negligible with high energy electrons; as a result, they would cancel each other.
\end{thebibliography}
\end{document}